\documentclass[preprint,12pt]{elsarticle}

\usepackage{amsmath,amssymb,amsfonts}
\usepackage{algorithmic}
\usepackage{graphicx}
\usepackage{textcomp}

\usepackage{subfigure}
\usepackage{color}
\usepackage{multirow}
\usepackage{bbm}
\usepackage{algorithm}

\journal{IEEE ACCESS}

\begin{document}

\begin{frontmatter}



\title{User-item matching for recommendation fairness\tnoteref{t1}}

\tnotetext[t1]{This article has been published on IEEE ACCESS, please cite as: Q. Dong*, S.-S. Xie, W.-J. Li, User-item matching for recommendation fairness, IEEE Access, 2021, 9: 130389-130398, DOI: 10.1109/ACCESS.2021.3113975.}


\author[wsc]{Qiang Dong\corref{cor}}
\cortext[cor]{Corresponding author.}
\ead{dongq@uestc.edu.cn}
\author[wsc]{Shuang-Shuang XIE}
\author[bdc]{Wen-Jun Li}

\address[wsc]{School of Computer Science and Engineering,
University of Electronic Science and Technology of China, Chengdu
611731, China}
\address[bdc]{School of Software and Service Outsourcing, Suzhou Vocational Institute of Industrial Technology, Suzhou 215004, China}

\begin{abstract}
As we all know, users and item-providers are two main parties of participants in recommender systems. However, most existing research efforts on recommendation were focused on better serving users and overlooked the purpose of item-providers.
This paper is devoted to improve the item exposure fairness for item-providers' objective, and keep the recommendation accuracy not decreased or even improved for users' objective.
We propose to set stock volume constraints on items, to be specific, limit the maximally allowable recommended times of an item to be proportional to the frequency of its being interacted in the past, which is validated to achieve superior item exposure fairness to common recommenders and thus mitigates the Matthew Effect on item popularity. With the two constraints of pre-existing recommendation length of users and our stock volumes of items, a heuristic strategy based on normalized scores and a Minimum Cost Maximum Flow (MCMF) based model are proposed to solve the optimal user-item matching problem, whose accuracy performances are even better than that of baseline algorithm in regular recommendation context, and in line with state-of-the-art enhancement of the baseline. What's more, our MCMF based strategy is parameter-free, while those counterpart algorithms have to resort to parameter traversal process to achieve their best performance.
\\
\\
\end{abstract}


\end{frontmatter}


\section{Introduction}
\label{introduction}

Even though the broad social and business acceptance of recommender systems has been achieved, a key underexplored dimension for further improvement is the usefulness of recommendations to the participants  \cite{over-specialization}. A recommender system usually serves two main parties of participants, the users and the item-providers \cite{Multistakeholder}, thus the usefulness of recommendations should be also two-fold. On one hand, recommender systems provide users with items of their latent interests. On the other hand, recommender systems should also help item-providers increase sales volume of items, especially the unpopular ones.

Common recommenders, such as collaborative filtering,  originally proposed to make accurate prediction of unseen user-item interactions, usually suffer the  \emph{popularity bias problem}, i.e., recommending a few popular items to a majority of users \cite{debias}. Although popular items are likely to match users' preferences and recommendations of them contribute to the predictive accuracy, users usually do not regard them as very useful recommendations because they are easily aware of these popular items somewhere else, for example from sales leaderboard, advertisements, or friends' conversations. Thus as a complement, the intra-list diversity \cite{ILD} is introduced to measure how well a recommender can widen a user's vision of items, usually by means of offering users less popular, unexpected but interesting items.

The popularity bias problem also hampers  the sales promotion of unpopular items, which is the main usefulness of recommendations for item-providers. Just like the intra-list diversity is used to measure how serious is the popularity bias from the point of view of individual users, we also need to measure this concentration problem for the purpose of item-providers. Therefore, The Gini coefficient is borrowed from the economic filed to quantify the balance degree in the numbers of recommended times of different items \cite{gini}, which is called \emph{exposure fairness} in this paper.

The most straightforward approach to improve the exposure fairness is, first generating a relatively large recommendation list with a classical model, and then performing post-hoc re-ranking on that recommendation list. Abdollahpouri \cite{aies} re-ranked the generated recommendation list by considering the item popularity, and Christoffel \emph{et al.} \cite{rp3} simply divided each item's recommendation score by its degree of popularity with an adjustable exponent on the shoulder, which greatly decreases the originally high recommendation scores of popular items and thus enhances the priority of unpopular items in the recommendation lists. Dong \emph{et al.} \cite{rank-agg} proposed to linearly aggregate the row- and column-ranking numbers of the recommendation score matrix obtained by some algorithm into the final recommendation ranking number, where the row and column of the matrix are corresponding to user and item, respectively. Mansoury \emph{et al.} \cite{Mansoury} introduced a general graph-based algorithm for improving item recommendation fairness. The algorithm iteratively finds items that are rarely recommended yet are high-quality and add them to the users’ final recommendation lists, which is done by solving the maximum flow problem on the recommendation bipartite graph. Item fairness in multi-round recommendation context also received research attentions. Patro \emph{et al.} \cite{Patro} focused on the item fairness issues arising out of incremental updates of the platform algorithms. They formulated an ILP based online optimization to ensure smooth transition of the exposure of items while guaranteeing a minimum utility for every user.
Ge \emph{et al.} \cite{Ge} explored the problem of long-term exposure fairness of items in dynamically changing groups of different popularity levels. They proposed a fairness-constrained reinforcement learning algorithm based on Constrained Markov Decision Process (CMDP), so that the model can dynamically adjust its recommendation policy.

Although the improved exposure fairness brought by these methods may be significant in terms of improved percentage, for example of more than 100\%, the improved absolute value is usually trivial, for example from 0.0378 to 0.0859 (see the values of exposure fairness of original P3 and RP3 algorithms on the Movielens data set in Table \ref{tab:heuristic-ml}). It is hard to say that this kind of improvement of exposure fairness will better serve the purpose of item-providers. Here what we should not ignore is that, L\"{u} \emph{et al.} \cite{congestion} improve the exposure fairness to a nontrivial absolute value, but with the cost of unacceptable loss of recommendation accuracy.

Our main task is to significantly improve the exposure fairness, and simultaneously keep the recommendation accuracy not decreased or even improved. The main contributions of this paper are:

\vspace{1ex}

(1) We propose to set stock volume constraints on items, to be specific, limit the maximally allowable recommended times of an item to be proportional to the frequency of its being interacted in the past, which is validated to achieve superior item exposure fairness to common recommenders and thus mitigates the Matthew Effect on item popularity.

\vspace{1ex}

(2) For the purpose of users, two heuristic user-item matching strategies are proposed to minimize the loss of recommendation accuracy brought by item stock volume constraints. Among them, the parameterized strategy is validated to achieve better recommendation accuracy than the baseline algorithm in regular recommendation scenario, and it has an advantage of relatively low time complexity.

\vspace{1ex}

(3) A Minimum Cost Maximum Flow based model is designed to solve the optimal user-item matching problem with constraints. The recommendation accuracy of this strategy is in line with state-of-the-art enhancement of the baseline algorithm, but it is parameter-free and get rid of parameter traversal process of its counterpart enhancements to achieve their best performance.

\section{Related works}

Generally speaking, the primary goal of a recommender system is to enhance
the engagement of users by providing them with items of potential interests. Although the usefulness of recommendation results is usually evaluated by accuracy measures (how accurate they are), the literature
has introduced different evaluation measures of the quality of recommendation results from different perspectives \cite{Silveira}. The most frequently-used and extensively-studied types of beyond-accuracy measures are coverage, novelty, diversity and fairness.

However recently, many bias types have been recently discovered and categorized into  bias in data,  bias in model and bias in results, respectively arising from three different stages of the recommendation feedback loop \cite{debias}, deteriorating the recommendation quality in terms of the above-mentioned beyond-accuracy aspects and challenging to achieve qualified recommendations. Among these bias types, popularity bias is the most
prominent one due to its highly adverse effects on beyond-accuracy
recommendation quality. In the literature, many methods have been developed for mitigating the bias problems in recommendations. These debiasing approaches can be divided into three main categories, pre-, in-, and post-processing methods, according to the three stages they participate in the recommendation process \cite{Boratto}. 


 
Since the data of user interactions are observational rather
than experimental, the imbalance in user-item interaction data becomes one of the main factors accounting for popularity bias. Pre-processing approaches
usually aim to reduce such inequalities by altering data
on which recommendation algorithms are trained. 
For example, Park \emph{et al.} \cite{Park} divide all the items into head group and tail group, where the head group consists of popular items with significantly larger amount of ratings than those in the tail group.
Recommendations for tail items are produced using only ratings
in the tail group, while those for head items are estimated using
all data. 
Jannach \emph{et al.} \cite{Jannach} present a practical popularity debiasing
technique that first creates synthetic user-item tuples where the
observed items are mostly unpopular and then utilizes them to
train algorithms. 
Chen \emph{et al.} \cite{Jiawei} derive a general learning framework that well summarizes most existing data debiasing strategies by specifying some parameters of the general framework. This provides a valuable opportunity to develop a universal solution for debiasing, e.g., by learning the debiasing parameters from data.

The in-processing approaches aim to modify the internal mechanisms of recommendation algorithms to simultaneously consider both popularity and relevance. This task is usually accomplished using specific constraints or conducting a joint optimization. 
For example, Abdollahpouri \emph{et al.} \cite{Controlling} propose an optimized variant of the well-known RankALS algorithm, which contributes to producing recommendation lists where predictive accuracy and intra-list item diversity are balanced. 
Hou \emph{et al.} \cite{Hou} present a framework
that first constructs the neighborhoods between the items based
on their popularity instead of the magnitude of their ratings; then, it eliminates some most popular ones to have a more balanced
common-neighbor similarity index. 
Boratto \emph{et al.} \cite{Boratto} propose an in-processing approach
that minimizes the biased correlation between user-item relevance and item popularity for fairly treating the items along the popularity tail.
Berbague \emph{et al.} \cite{Berbague} propose a solution to balance between the recommendation accuracy and coverage by making an overlapped clustering, where each user is assigned to a main cluster from which he gets his recommendations and to secondary clusters as a candidate neighbor.

The post-processing techniques usually aim to re-rank a recommendation
list that has already been generated or create a new
one following some specific constraints. These are the most utilized
approaches for mitigating popularity bias since they can be easily
applied to the output of any recommendation algorithm, which
is also why we focus on developing a post-processing method to
counteract potential popularity bias in this study. 
For example, Abdollahpouri \emph{et al.} \cite{Popularity-aware} introduce an approach that first calculates weight scores for items based on their popularity and then utilizes them to punish popular items during re-ranking recommended item lists. 
Likewise, Yalcin and Bilge \cite{Yalcin} follow a similar strategy and presents two robust popularity debiasing methods for recommending to groups of users
rather than individuals. 
Abdollahpouri \emph{et al.} \cite{Managing} also present the xQuad algorithm that is an enhanced re-ranking approach and helps balance the trade-off between long-tail item coverage and ranking accuracy more robustly.
Wu \emph{et al.} \cite{Wu} exploite a balance factor to adjust the influence of a personalized ranking vector and a unified non-personalized ranking vector based on PageRank. By this, it can reduce the impact of item popularity on recommendations and then generate more diverse and novel recommendations to users.

As seen from the presented literature, many recent studies consider
biases towards popular items in the recommendations and
try to deal with this issue for achieving more qualified recommendations.
However, as emphasized in the previous section, the popularity
of an item does not always mean that individuals strongly
desire it. Therefore, more comprehensive analyses are required to
investigate potential bias towards blockbuster items and develop
novel practical methods to mitigate its adverse effects on recommendations.

\section{Problem description and notations}


\subsection{Notations}

In an implicit rating based recommender system  with $m$ users and $n$ items, $u$ or $u_i$ ($1 \leq i \leq m$) is used to denote a user, $v$ or $v_j$ an item ($1 \leq j \leq n$), and the adjacency matrix $A = (a_{i,j})_{m\times n}$ represents the historical user-item interaction records.
The matrix element $a_{i,j}$ is 1 if there exists an observed user-item interaction, indicating that user $u_i$ declared explicitly his/her preference on item $v_j$ in the past, and the element is 0 if otherwise. The sum of row $i$ is called the degree of user $u_i$, denoted by $deg(u_i)$, which is an indicator of the user's activity level in the system. The sum of column $j$ is called the degree of item $v_j$, denoted by $deg(v_j)$, which represents the item's popularity degree among all users.

Given the user-item adjacency matrix $A$ as input, a recommendation algorithm will output the user-item score matrix $S=(s_{i,j})_{m\times n}$, where the element $s_{i,j}$ is the predicted score of user $u_i$'s preference on item $v_j$.
The recommendation list of each user is constituted by the items with top $l$ recommendation scores. Of course, the interacted items are excluded from the recommendation list.
The final recommendation results of all users can also be represented by the recommendation matrix $R = (r_{i,j})_{m\times n}$, where $r_{i,j} = 1$ if item $v_j$ is recommended to user $u_i$, otherwise zero. Clearly, the sum of every row of matrix $R$ must equal the \emph{recommendation length} $l$.
All the notations used in this paper are listed in Table \ref{tab:notations}, some of which will be defined in the remaining part.

\begin{table}
\small
\caption{The notations used in this paper.}
\label{tab:notations}
\begin{center}
\begin{tabular}[b]{lp{10cm} }
\hline
Notation&               Definition\\
\hline
\vspace{1ex}
$m$ &                   Number of users\\
\vspace{1ex}
$n$ &                   Number of items\\
\vspace{1ex}
$i$, $j$, $k$ &	        Counter\\
\vspace{1ex}
$u$, $u_i$ &            Some user\\
\vspace{1ex}
$v$, $v_j$ &            Some item\\
\vspace{1ex}
$A=(a_{ij})_{m\times n}$&	        User-item adjacent matrix\\
\vspace{1ex}
$deg(u)$, $deg(u_i)$ &	Degree of some user\\
\vspace{1ex}
$deg(v)$, $deg(v_j)$ &	Degree of some item\\
\vspace{1ex}
$S=(s_{ij})_{m\times n}$	&       Matrix of recommendation scores, but $s_{ij}$ is set to be 0 if \vspace{1ex} $a_{ij} = 1$.\\
\vspace{1ex}
$S^N=(s^N_{ij})_{m\times n}$	&       Row-normalized matrix of $S$\\
\vspace{1ex}
$R=(r_{ij})_{m\times n}$ &    	Matrix of recommendation results, where $r_{ij} =1$ if item $v_j$ is recommended to user $u_i$, otherwise zero\\
\vspace{1ex}

$l$ &               	Length of recommendation lists, or initial vacancy number of the recommendation list of every user\\

\vspace{1ex}
$L$ &	                Vector of remaining vacancy numbers of the recommendation lists of $m$ users\\

\vspace{1ex}
$L(i)$	&         Remaining vacancy number of the recommendation list of user $u_i$\\
\vspace{1ex}
$q_j$ &	          Initial stock volume of item $v_j$\\
\vspace{1ex}
$Q$ &	          Vector of remaining stock volumes of $n$ items\\
\vspace{1ex}
$Q(j)$ &	      Remaining stock volume of item $v_j$\\
$\mathbbm{1}$ &	  Vector consisting of all one elements\\

\hline
\end{tabular}
\end{center}
\end{table}

\subsection{Item stock volume constraints}

The recommendation process is essentially selecting a subset of items to fill into the vacancies of the recommendation list of every user. Given user number $m$ and  recommendation length $l$, the total number of vacancies in all users' recommendation lists is $m* l$. In order to limit the recommendation frequency of popular items and promote the exposure of less popular items, we set the maximally allowed recommendation frequency of a specific item $v_j$, called \emph{stock volume} $q_j$, to be proportional to its degree as follows,

\[{q_j} = \left\lceil {m * l * \frac{{\deg \left( {{v_j}} \right)}}{{\sum_{k = 1}^n {\deg \left( {{v_k}} \right)} }}} \right\rceil . \]

\noindent The value is rounded up to an integer, because the total stock volume of all items must be no less than the vacancy number in the recommendation lists of all users.

By constraining the stock volume of every item, the relative popularity of different items will remain almost unchanged before and after recommendation process, and the Matthew effect on item popularity will be prevented from being aggravated, with the hypothesis that all the item recommendations shares the same conversion rate. Of course if the varying conversion rate is available, we can incorporate it as the weight of recommendation in the following fairness measure.

Next let us justify the feasibility of this constraint on item stock volume in a real-world recommendation context. In the service recommendation scenarios like dining, accommodation, fitness, haircuts, massages, medical services and so on, stock constraint is a special factor that decides how many customers can receive a service with an assured level of quality. For example, a restaurant often has a constraint on the number of customers who can be served during the dining hours. If too many customers arrive at a restaurant, their dining experience will be unpleasant or in the worst case, some customers will be very disappointed  \cite{FAST}.


\subsection{Problem formalization}

In the regular recommendation scenario, there exists only constraint on the user side (the recommendation length) and no constraint on the item side, thus the recommendation lists of different users are generated independently of each other. That is to say, recommending which items to one specific user has nothing to do with the recommendation list of other users. In this paper, the item stock volume is introduced as one constraint on the item side. These two constraints together break the independency of different recommendation lists. The arising problem is, in order to achieve the best recommendation quality, how to match the items with limited stock volume into different users' recommendation lists of fixed length? For example, some regular algorithm is supposed to recommend a specific item to many users, but the stock volume of this item is fewer than the amount of these users.

This is an optimal user-item matching problem with constraints, where the optimization  objective is undoubtedly the overall recommendation accuracy, and the two constraints are user's recommendation length and item's stock volume. Since user's future feedback is unknown during the recommendation generation phase, the most simple and intuitive alternative for the recommendation accuracy is the sum of recommendation scores of all matched user-item pairs in the recommendation lists.

Empirical results \cite{qiu} show that there exists a strong positive correlation between the recommendation scores and the user degrees (item degrees, respectively). That is to say, almost all the users are regarded to prefer popular items to unpopular ones by common recommenders, and the recommendation scores on popular items of large-degree users is usually much larger than that of small-degree users. Therefore, following the above optimization objective, the large-degree users will almost run out the stocks of popular items, and the small-degree users who also prefer popular items have to be recommended less preferable unpopular items. Since the small-degree users are the absolute majority in the system, this will cause unacceptable loss of recommendation accuracy.

In order to eliminate the influence of user degree on the recommendation scores, we propose to use the user-normalized scores instead of original scores for the user-item matching priority. Specifically, the user-normalized score of every item  is defined by the original score divided by the sum of all items' scores of the same target user,

\[s^N_{ij} = \frac{{{s_{ij}}}}{{\sum_{ k =1 }^n {{s_{ik}}} }}.\]

\noindent By means of the score normalization process, every user holds the same amount of recommendation stakes to be assigned to all the candidate items (the sum of normalized scores is one unit for every user). However, the distribution of stakes among items varies from small- to large-degree users. The stakes of small-degree users are concentrated on a small number of popular items, such that although small-degree users enjoy higher priority in the user-item matching process, the number of priority times is very few for each small-degree user. In this way, most stocks of popular items will be exhausted by small-degree users, and the large-degree users have to be recommended less popular items.

In summary, the optimization problem of user-item matching with constraints is formulated as

\begin{equation*}
\begin{aligned}
\rm{max} & \sum\limits_{1 \le i \le m \atop 1 \le j \le n} {{r_{ij}}{ s^N_{ij}    }}\\
\rm{s.t.} & \sum_{j =1}^n {{r_{ij}}}  = l, & 1 \le i \le m\\
& \sum_{i=1}^m {{r_{ij}}}  \le {q_j}, & 1 \le j \le n.
\end{aligned}
\end{equation*}

\section{User-item matching strategy}

To solve the optimal user-item matching problem with constraints, we propose two kinds of priority strategies, one is the simplest greedy strategy and the other one is the elaborate Minimum Cost Maximum Flow based strategy.

\subsection{The greedy strategy}

The essence of our greedy user-item matching strategy is "Largest Normalized-Score First". For this strategy, we sort all the non-interacted user-item pairs $(u_i,v_j)$ in descending order of normalized scores, and check each pair one by one in this order. For a specific user-item pair, if there is some vacancy in the corresponding user's recommendation list and the corresponding item's stock volume is not exhausted, we fill the item into the recommendation list of the user, and decrease the corresponding vacancy number and stock volume by one. Once the recommendation lists of all users are fully occupied by items, we stop the above checking process. Algorithm \ref{alg:Score} presents the detailed steps of this greedy strategy.

\begin{algorithm}[htb]
\caption{Largest-Normalized-Score-First}
\label{alg:Score}
\begin{algorithmic}[1] 
\REQUIRE ~~\\ 
user-item adjacent matrix $A$, normalized score matrix $S^N$, Vector $L$ of users' initial numbers of recommendation vacancy, vector $Q$ of items' initial stock volume.

\ENSURE ~~\\ 
recommendation matrix $R$ such that $R \times \mathbbm{1} = L$ and $R^T \times \mathbbm{1}\leq Q$.

\STATE Initialize $R$ as zero matrix.

\STATE Sort the $m\times n$ elements of row-normalized score matrix $S^N$ in descending order, and assume that the $k$-th largest element in this order is originally in row $i_k$, column $j_k$ of matrix $S^N$.

\FOR{$k = 1$ to $m\times n$}
    \IF[the recommendation lists of all users are fully occupied*/]{$L = \vec{0}$}
        \STATE break;
    \ENDIF

    \IF[there is available vacancy in the recommendation list of the user and there is remaining stock volume for the item*/]{$L(i_k) > 0$ and $Q(j_k) > 0$}
        \STATE $R(i_k, j_k) = 1$; \COMMENT{fill the item in the list of the user*/}        \STATE $L(i_k) = L(i_k) - 1$; \COMMENT{decrease the remaining vacancy number of the user*/}
        \STATE $Q(j_k) = Q(j_k) - 1$; \COMMENT{decrease the remaining stock volume of the item*/}
    \ENDIF
\ENDFOR
\RETURN $R$;

\end{algorithmic}
\end{algorithm}

\subsection{The MCMF-based strategy}

As one important contribution of this paper, we build a Minimum Cost Maximum Flow (MCMF) model to solve the optimal user-item matching problem with constraints. The MCMF problem is a well-known network flow problem, which finds various applications in the fields of transportation, logistics, telecommunication, network design, resource planning, scheduling, and many other industries \cite{optics}. Next we give a brief review on the MCMF problem. The terminology follows from the reference \cite{book}.

A network is a directed graph $G = (V, E)$ with a source node $s$  and a sink node  $t$. Each directed edge $(u,v) \in E$  is associated with two constants, the capacity  $cap(u,v)$ indicating the upper bound of the flow $f(u,v)$  allowed on the edge, and the cost per unit flow on the edge, denoted by  $cost(u,v)$. Clearly, the capacity and the cost of an edge are positive values. The value of a network flow $f$ is defined as
$value\left( f \right) = \sum\nolimits_{\left( {s,w} \right) \in E} {f\left( {s,w} \right)}  - \sum\nolimits_{\left( {w,s} \right) \in E} {f\left( {w,s} \right)} $, and the cost of flow $f$ is  ${\mathop{\rm cost}\nolimits} \left( f \right) = \sum\nolimits_{\left( {s,w} \right) \in E} {f\left( {s,w} \right) \times {\mathop{\rm cost}\nolimits} \left( {s,w} \right)}  - \sum\nolimits_{\left( {w,s} \right) \in E} {f\left( {w,s} \right) \times {\mathop{\rm cost}\nolimits} \left( {w,s} \right)} $. A Minimum Cost Maximum Flow of a network $G = (V, E)$ is a maximum flow with the smallest possible cost.

To relate the optimization problem of user-item matching to the MCMF problem, the most important work is to construct a flow network $G = (V, E)$ to model the optimization objective and the constraints of the user-item matching problem, which is defined as follows. The node set is

\[V = \left\{ {s,t} \right\} \cup {\left\{ {{u_i}} \right\}_{1 \le i \le m}} \cup {\left\{ {{v_j}} \right\}_{1 \le j \le n}}\]

\noindent where $u_i$ and $v_j$ represent users and items of the recommender system, consistent with the aforementioned notations.

The directed edge set is

\begin{equation*}
\begin{aligned}
E = {\left\{ {\left( {s,{u_i}} \right)} \right\}_{1 \le i \le m}} & \cup {\left\{ {\left( {{v_j},t} \right)} \right\}_{1 \le j \le n}}\\
& \cup {\left\{ {\left( {{u_i},{v_j}} \right)} \right\}_{a_{i,j}=0,\atop 1 \le i \le m, 1 \le j \le n}}
\end{aligned}
\end{equation*}

\noindent In other words, there is a directed edge from the source node to each user node, a directed edge from each item node to the sink node, and a directed edge between every user-item pair without interaction in the past.

The capacity on each directed edge is defined as

\begin{equation*}
\begin{aligned}
& cap\left( {s,{u_i}} \right) = l,cap\left( {{u_i},{v_j}} \right) = 1, \\
& cap\left( {{v_j},t} \right) = {q_j}, 1 \le i \le m,1 \le j \le n.
\end{aligned}
\end{equation*}

The edge capacity between every user-item pair is set to be 1, indicating the user-item matching rule that each item can occupy at most one vacancy of the recommendation list of a specific user. The edge capacity from the source node to each user node is defined by the length of recommendation lists, and the edge capacity from each item node to the sink node is defined by the stock volume of the item, corresponding to the two constraints of the optimization function.

The cost on each directed edge is defined as

\begin{equation*}
\begin{aligned}
& {\mathop{cost}\nolimits} \left( {{u_i},{v_j}} \right) = 100 * (1 -  \lceil s^N_{i,j} \rceil),\\
& {\mathop{cost}\nolimits} \left( {s,{u_i}} \right) = {\mathop{cost}\nolimits} \left( {{v_j},t} \right) = 0, 1 \le i \le m,1 \le j \le n,
\end{aligned}
\end{equation*}

\noindent where the normalized score $s^N_{i,j}$ is rounded up to two decimals, such that the values of capacity and cost of this network are all positive integers.

After construction of the directed flow graph, we relate the original optimization problem to the MCMF problem. One benefit of making the connection between the user-item matching problem and the MCMF problem is that it provides an approach for taking advantage of existing works that have already been done on finding the optimal solution. The MCMF problem has been thoroughly studied and many efficient MCMF algorithms are available in the literature \cite{book,algorithm-book}.

The value of the maximum flow of this network is clearly $m*l$, since we get the initial stock volumes of items by rounding up the values. The objective of minimizing the cost of the maximum flow is essentially the optimization objective of maximizing the sum of normalized scores of matched user-item pairs.

To compute a minimum cost maximum flow in graph $G = (V, E)$ from $s$ to $t$, we employ the Capacity Scaling algorithm and the MCMF problem can be solved in polynomial time \cite{book}. The solution of the MCMF problem yields the result of the optimal user-item matching problem. The recommendation list of a target user is consisted of items with nonzero flow on the corresponding user-item edges in the optimal solution of the MCMF problem.

\section{Performance evaluation}

\subsection{Data sets, evaluation measures and baseline algorithms}

Two benchmark data sets are employed to evaluate the performance of recommendation algorithms, namely, Movielens and Netflix. Both of them are movie rating data set, where users rate their watched movies (rephrased as items in this paper) with an explicit integer scores from 1 to 5. For each data set, we use only the ratings no less than 3 to construct the nonzero elements of adjacent matrix of user-item interactions. Table \ref{tab:dataset} summarizes the statistical features of the two data sets,  where the sparsity is the proportion of nonzero elements to the total number of elements of adjacent matrix. To evaluate the offline performance of different recommendation algorithms, each data set is temporally partitioned into two subsets: the training set containing early 80\% of the nonzero elements and the probe set later 20\% of the nonzero elements. The training set is treated as known information to make recommendation and the probe set is used to test the accuracy performance of the recommendation results.

\begin{table}
\caption{The basic statistics of two real-world networks used in this paper, including the number of users, item and links, and the sparsity.}
\label{tab:dataset}
\begin{center}
\begin{tabular}[b]{lllll }
\hline
data set&	\#users&      \#items&	\#links&	sparsity\\
\hline
Movielens&	6000&        3,600&	800,000&	3.8\%\\

Netflix&	9500&       14,000&	1,700,000&	1.2\%\\
\hline
\end{tabular}
\end{center}
\end{table}

The most simple recommendation fairness measure is the \emph{aggregate diversity} \cite{aggregateD1,aggregateD2} (also known as coverage \cite{xiang}), which is defined by the fraction of items recommended to at least one user to the total number of items. This intuitive measure may not be very robust, since the contribution to it of an item that has been recommended just once is equal to that of other item recommended a thousand times. To solve this problem, Fleder and Hosanagar \cite{gini} proposed a better alternative by using the Gini coefficient to measure the balance in the numbers of recommended times of different items,

\[G = 1 - \frac{1}{{n - 1}}\sum\limits_{k = 1}^n {\left( {2k - n - 1} \right)p\left( {{i_k}|R} \right)} \]

\noindent where $p\left( {{i_k}|R} \right)$ is the probability of the $k$-th least recommended item being drawn from the recommendation lists generated by a recommender system. In order to be in accordance with other metrics for which higher value is better, the complement of the standard definition of Gini coefficient is used in this paper.

Besides the above two item-provider oriented measures, we use \emph{Precision} to evaluate the quality of recommendation results for the purpose of users, which is defined as the fraction of accurately recommended items to the length of recommendation lists \cite{xiang}.

In this paper, we use the P3 algorithm \cite{P3} (also known as NBI \cite{NBI} or ProbS \cite{pnas}  algorithm) as the baseline, and several well-known P3-enhanced algorithms as comparing counterparts, including the RP3, HHP, RAP3, PD, BD and BHC algorithms \cite{rp3,rank-agg,pnas,lv-pd,Nie,bhc}. The interested reader is referred to a survey article for comprehensive review \cite{zeng-an}.

Our proposed user-item matching strategy with constraints are generic post-processing methods which can be used to improve any algorithms. Among many classic recommendation models, the reasons of selecting P3 algorithm as the baseline are as follows. First, P3 is intrinsically a hybrid form of user-based and item-based collaborative filtering with diffusion-based similarity \cite{zeng-an}. Second, the P3 algorithm does not require any pre-specified parameter, such as the neighborhood size in the $k$-nearest neighbor collaborative filtering. Third, the P3 algorithm has a perfect physical interpretation, since it is analogous to a mass diffusion process on the user-item network \cite{pnas}. Finally, the spreading representation of P3 in sparse networks is computationally more efficient than the traditional matrix-based representation of collaborative filtering methods \cite{zeng-an}.

\subsection{Performance of greedy strategy and its improvement}

According to Table \ref{tab:heuristic-ml} and Table \ref{tab:heuristic-nl}, the precision value of the greedy strategy is about $90\%$ of that of original P3 algorithm on the Movielens data set, and $77\%$ for Netflix. This percentage of accuracy loss is definitely not acceptable for practical applications. The subsequent question is that, is it possible to regain $100\%$ of accuracy value of the regular recommendation scenario?

\begin{table*}
\footnotesize
\caption{Performance comparison of P3 algorithm in the constrained recommendation scenario and the performance of several P3-enhanced algorithms in regular recommendation scenario on the Movielens data set.}
\label{tab:heuristic-ml}
\begin{center}
\begin{tabular}[b]{p{1.5cm}p{5.5cm}p{1.3cm}p{1.5cm}p{1.5cm} }
\hline
Algorithm&	Recommendation Settings&     Precision&	Aggregate diversity&	Exposure fairness\\
\hline

\multirow{5}*{P3}& Regular&	0.1949&		0.0991&		0.0378\\

~& Constraint, original greedy strategy&	0.1757&		0.6493&			\textbf{0.2993}\\
~& Constraint, greedy strategy, $\theta = 0.9$&	0.1989&		\textbf{0.6693}&			\textbf{0.3002}\\
~&	Constraint, MCMF strategy &	0.2115&	0.5689&		\textbf{0.2973}\\
\hline
RP3&	Regular, $\lambda=0.6$&	\textbf{0.2289}&	0.3485&		0.0859\\
RAP3&	Regular, $\lambda=0.6$&	0.2082&	0.2945&		0.074\\
HHP&	Regular, $\lambda=0.3$&	0.2238&	0.2529&		0.0715\\
BHC&	Regular, $\lambda=0.8$&	0.218&	0.3115&		0.0914\\
BD&	Regular, $\lambda=0.7$&	\textbf{0.2297}&	0.3507&		0.0932\\
PD&	Regular, $\lambda= -0.7$&	\textbf{0.2283}&	0.332&		0.0828\\

\hline
\end{tabular}
\end{center}
\end{table*}

\begin{table*}
\footnotesize
\caption{Performance comparison of P3 algorithm in the constrained recommendation scenario and the performance of several P3-enhanced algorithms in regular recommendation scenario on the Netflix data set.}
\label{tab:heuristic-nl}
\begin{center}
\begin{tabular}[b]{p{1.5cm}p{5.5cm}p{1.3cm}p{1.5cm}p{1.5cm} }
\hline
Algorithm&	Recommendation Settings&     Precision&	Aggregate diversity&	Exposure fairness\\
\hline
\multirow{5}*{P3}&	Regular&	0.136&	0.0493&		0.0068\\

~&	Constraint, original greedy strategy&	0.1042&	0.5278&		\textbf{0.1204}\\
~&	Constraint, greedy strategy, $\theta = 0.9$&	0.1418&	0.5566&		\textbf{0.124}\\
~&	Constraint, MCMF strategy&	0.1473&	0.5956&		\textbf{0.1232}\\
\hline
RP3&	Regular, $\lambda=0.5$&	0.1533&	0.652&		0.0414\\
RAP3&	Regular, $\lambda=0.2$&	0.1542&	\textbf{0.7699}&		0.0676\\
HHP&	Regular, $\lambda=0.2$&	\textbf{0.156}&	0.3311&		0.0217\\
BHC&	Regular, $\lambda=0.8$&	0.1512&	0.2673&		0.017\\
BD&	Regular, $\lambda=0.7$&	\textbf{0.1596}&	0.5965&		0.0478\\
PD&	Regular, $\lambda= -0.7$&	0.1494&	0.6674&		0.0351\\
\hline
\end{tabular}
\end{center}
\end{table*}

Recall that the greedy strategy degrades the priority of large-degree users in the user-item matching process by dividing the original scores with the sum of recommendation scores of all the items of the same target user. Then a natural question arises, is the strength of this priority degradation optimal? To answer this question, we replace the constant exponent 1 of the denominator with an adjustable parameter $\theta$, that is to say, the original recommendation scores are divided by the sum of recommendation scores of the target user with exponent $\theta$ on the shoulder,

\[s^\theta_{ij} = \frac{{{s_{ij}}}}{(\sum_{ k =1 }^n {{s_{ik}}} )^\theta}.\]

\noindent By traversing this parameter $\theta$ from 0 to 1 to obtain the best recommendation accuracy, we enhance the original greedy strategy to a parameterized version. For a specific value of parameter $\theta$, the user-item matching process is similar to the original greedy strategy.

%
%
%

According to Table \ref{tab:heuristic-ml} and Table \ref{tab:heuristic-nl} , the best precision value of the parameterized greedy strategy is larger than that of original P3 algorithm. What is more, the aggregate diversity is almost 7 times of the original value on the Movielens data set, and more than 11 times for Netflix; the exposure fairness is almost 8 times  of the original value on the Movielens data set, and more than 18 times for Netflix.

\subsection{Performance of the MCMF strategy}

Table \ref{tab:heuristic-ml} and Table \ref{tab:heuristic-nl} also present detailed  performance comparison between the P3 algorithm with the MCMF strategy, and six state-of-the-art P3-enhanced algorithms in the regular recommendation scenario on the Movielens and Netflix data sets, where the values of their intrinsic parameters are set to be associated with the best recommendation precision.
For the P3 algorithm in the constrained recommendation scenario, the MCMF model achieves the best accuracy performance among all the user-item matching strategies (even better than the parameterized greedy strategy), thus we call it the \emph{P3-MCMF algorithm} as a whole for later discussion.

Compared with six state-of-the-art P3-enhanced algorithms in the regular recommendation scenario, the P3-MCMF algorithm achieves the recommendation precision of more than $90\%$ of the best value of all the enhanced algorithms; its exposure fairness value is more than three times (two times, respectively) of the best values of all the enhanced algorithms on the Movielens (Netflix, respectively) data set;  As for the aggregate diversity, the P3-MCMF algorithm achieves much better performance on Movielens and a little worse performance on Netflix, compared with the best values of all enhanced algorithms. Since the aggregate diversity is a very coarse measure of recommendation fairness, its trivial loss does not matter for practical applications.

Recall that these six counterpart algorithms are already enhanced versions of the P3 algorithm, with much better recommendation performance. Take the RP3 algorithm as an example. Compared with the P3 algorithm on Movielens, the precision is improved about $17\%$, the aggregate diversity is improved about $250\%$, and the exposure fairness is improved more than $100\%$. Although the six P3-enhanced algorithms are inspired by different motivations, they have similar recommendation performance in regular recommendation scenario (for example the precision is about 0.22 and the exposure fairness is below 0.1 on the Movielens data set), which indicates that further performance improvement will be a very difficult task. However, our MCMF strategy successfully cracks this problem.

The most important significance of our P3-MCMF algorithm is that, it is parameter-free and thus achieves this superior performance without the time cost of parameter optimization, while all the above existing enhanced algorithms have to traverse their intrinsic parameter to get the best performance.

\begin{figure*}[!htb]
\centering
  \subfigure[RP3]{
    \includegraphics[width=2.4in]{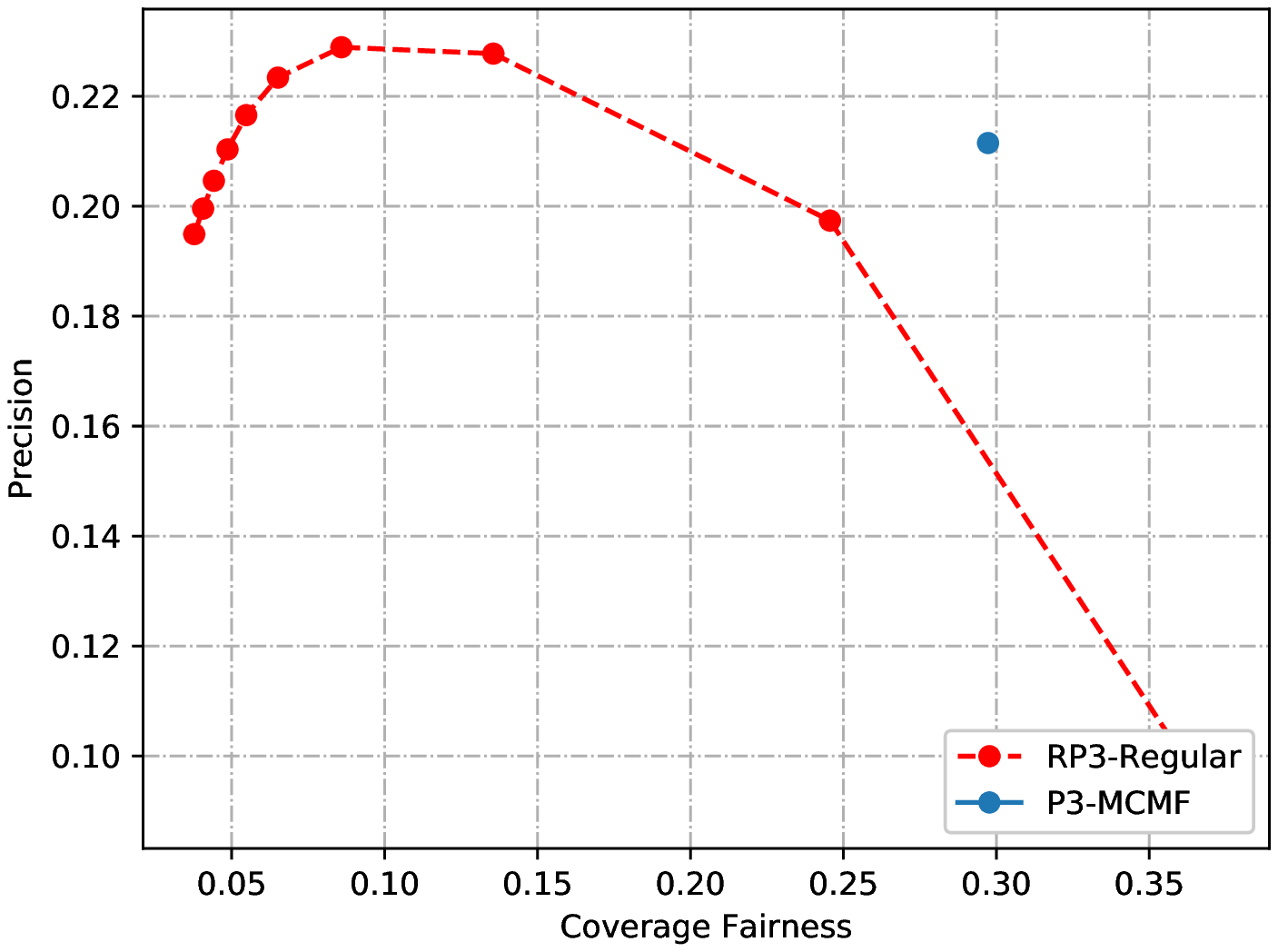}}
  \subfigure[HHP]{
    \includegraphics[width=2.4in]{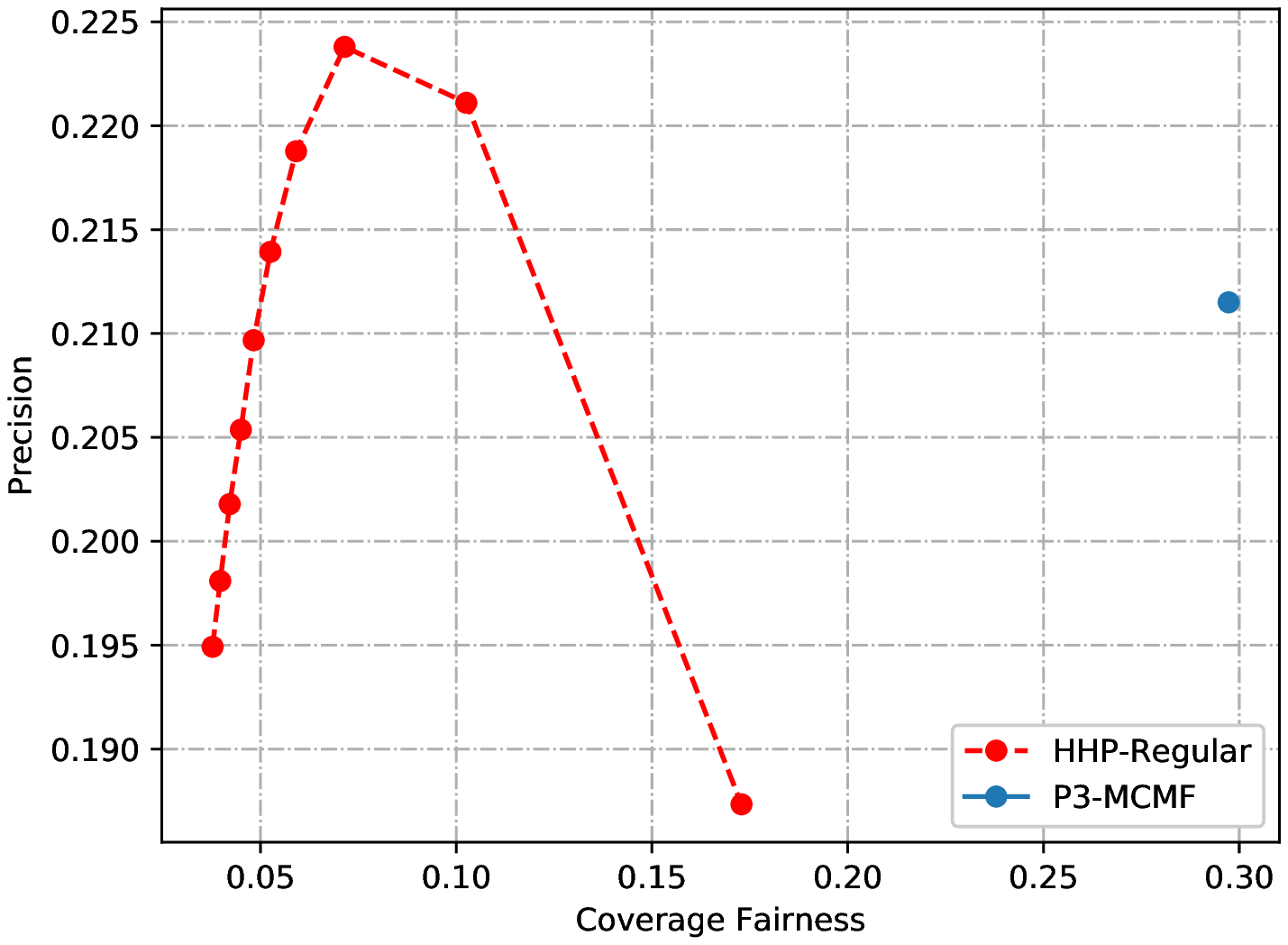}}
  \subfigure[BHC]{
    \includegraphics[width=2.4in]{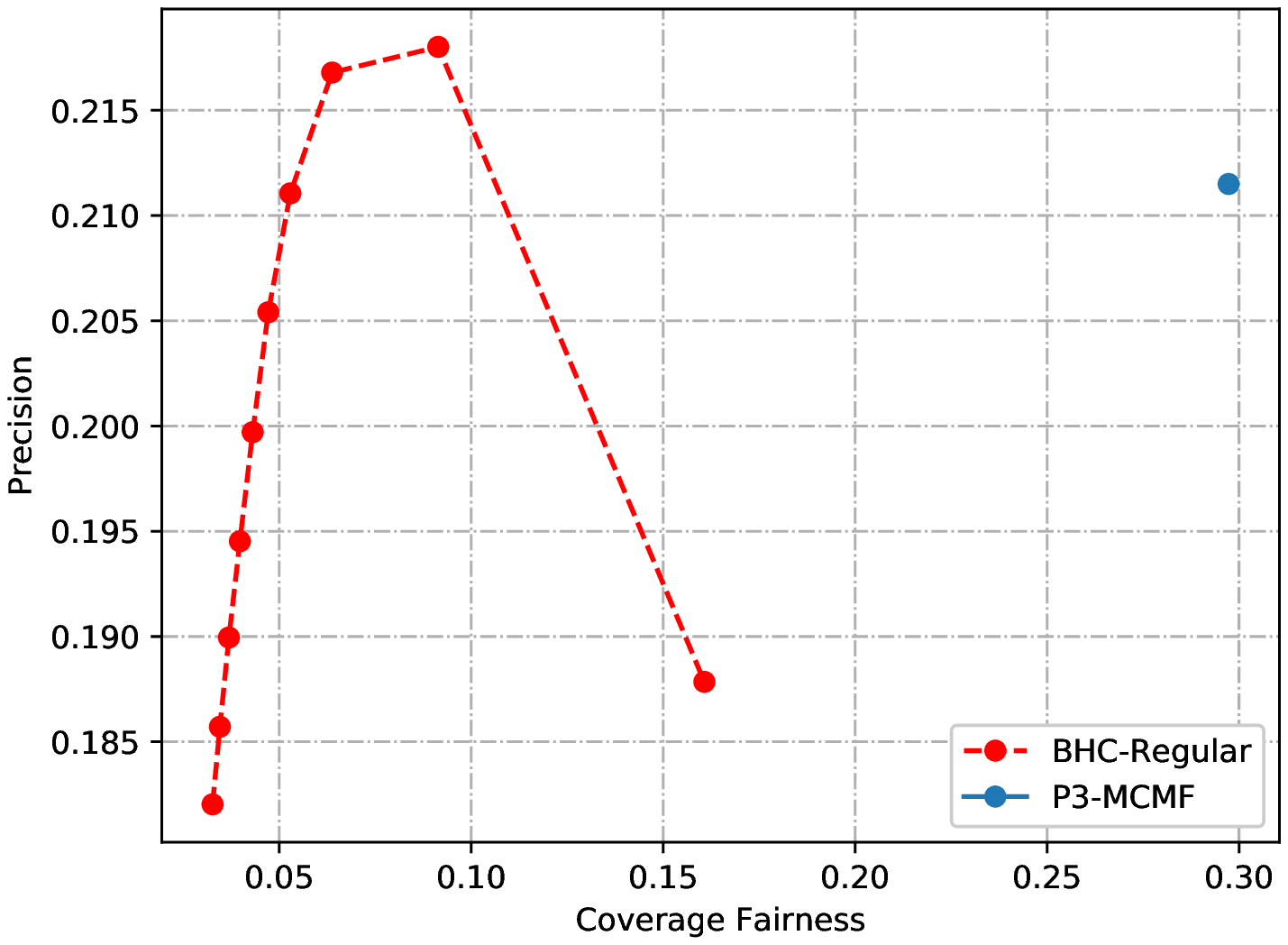}}
  \subfigure[BD]{
    \includegraphics[width=2.4in]{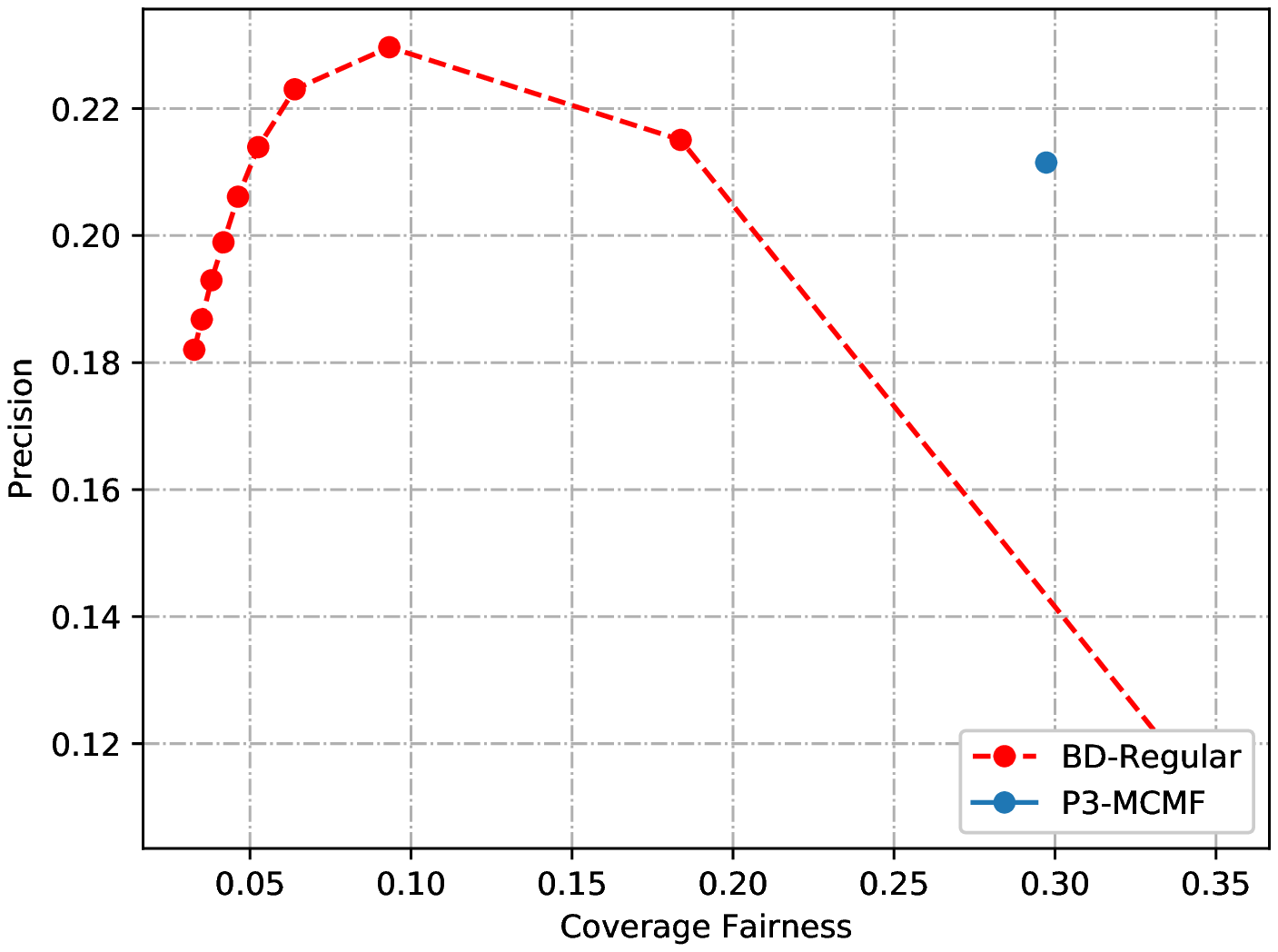}}
  \subfigure[PD]{
    \includegraphics[width=2.4in]{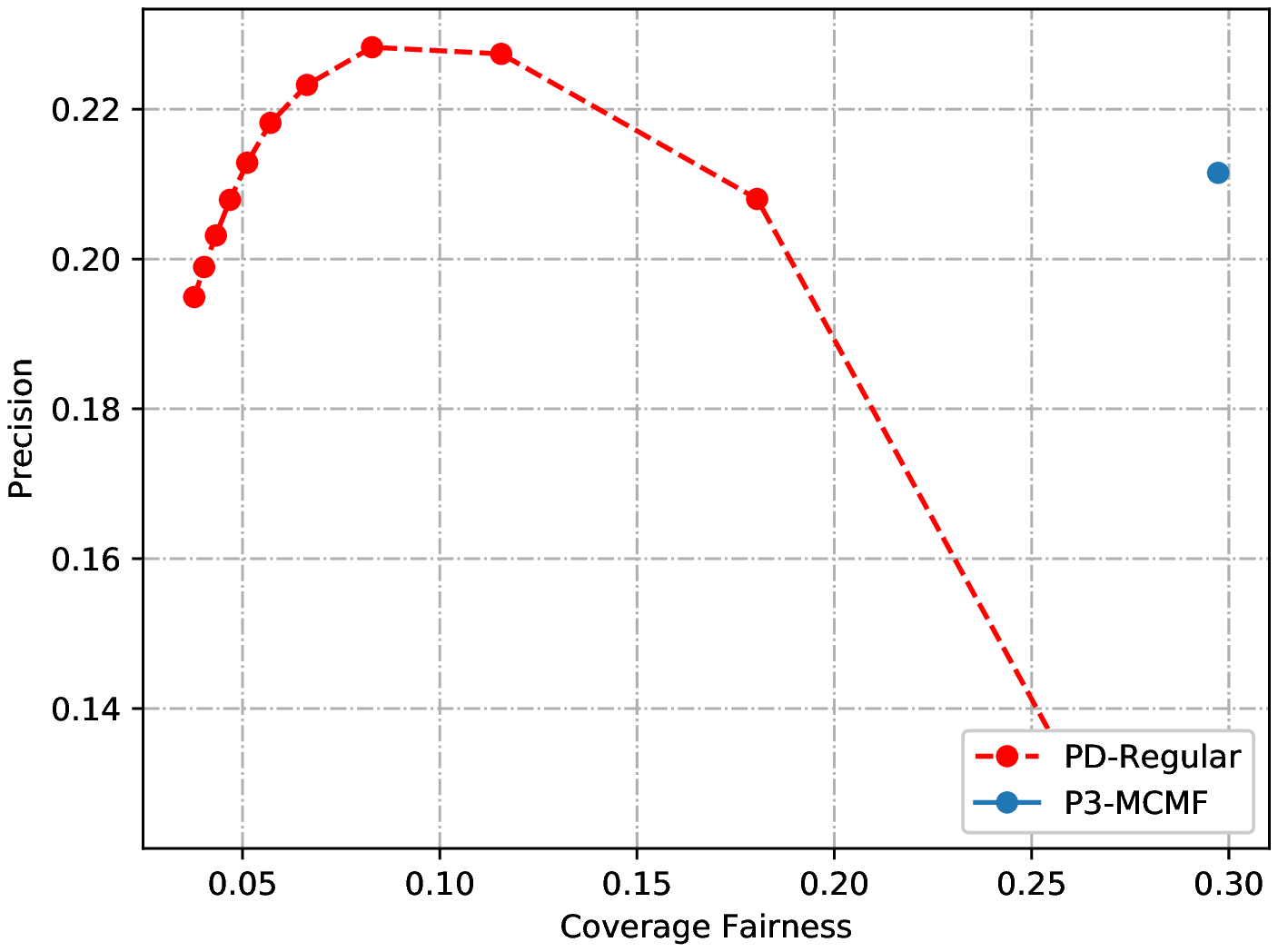}}
  \subfigure[RAP3]{
    \includegraphics[width=2.4in]{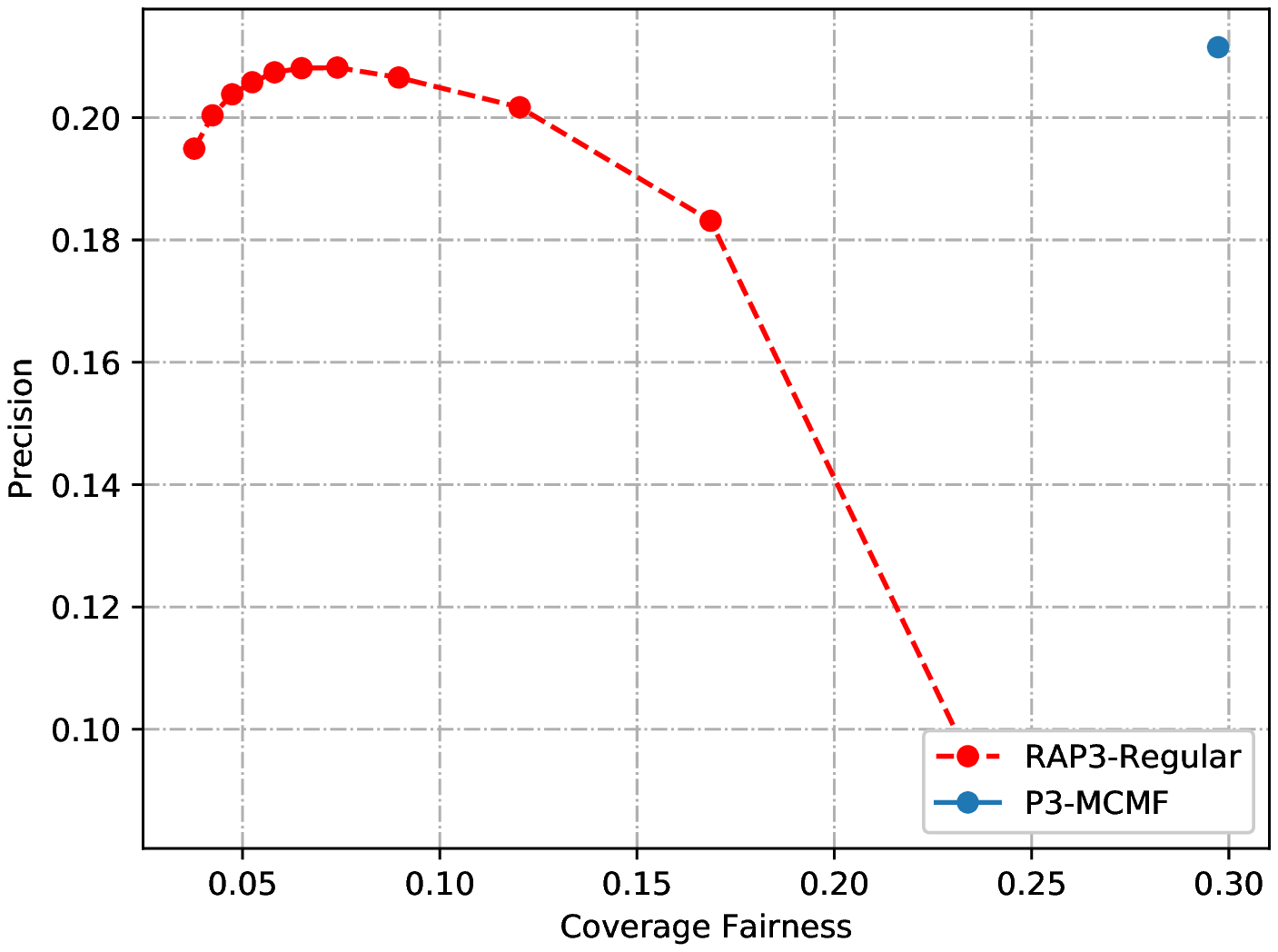}}
  \caption{The precision against the exposure fairness of six typical P3-enhanced algorithms (the red dots) with the change of their intrinsic parameter $\lambda$, compared with the counterparts of our P3-MCMF algorithm (the blue dot) on the Movielens data set. The length of recommendation lists is 20.}
\label{fig:further-1}
\end{figure*}

After the above performance comparison, a natural question is raised, can we adjust the parameter of some typical P3-enhanced algorithms, to get the similar exposure fairness value while sacrificing the accuracy to the same level? To answer this question, we plot the precision against the exposure fairness of six typical P3-enhanced algorithms, compared with the counterparts of our P3-MCMF algorithm on the Movielens data set in Fig. \ref{fig:further-1}. Each red dot is corresponding to a specific value of the intrinsic parameter $\lambda$ of the enhanced algorithm, from 0.1 to 1.0 with step length of 0.1, and the blue dot is for the performance of the P3-MCMF algorithm.

From Fig. \ref{fig:further-1} we can see that, the P3-MCMF algorithm regains more than $90\%$ of the best precision values of typical P3-enhanced algorithms. Even if we let the precision of P3-improved algorithms aligned with that of the P3-MCMF algorithm, the exposure fairness value of the former is still worse than that of the latter, and vice versa. That is to say, the performance of the parameter-free P3-MCMF algorithm is always better than those of the parameterized P3-enhanced algorithms. What is more, the parameter-free characteristic of the P3-MCMF algorithm is a tremendous advantage in the practical applications.

\subsection{Analysis of time complexity}	

The greedy user-item matching strategy consists of two steps: the sorting step of $mn$ elements of normalized matrix $S^N$ whose time complexity is $O(mn\log (mn))$, and the one by one checking step of $mn$ user-item pairs whose time complexity is $O(mn)$, thus the overall time complexity of greedy strategy is $O(mn\log (mn))$. According to \cite{book}, the time complexity of our MCMF strategy is $O(m^2n^2\log m\log (mn))$.

In summary, each of our two user-item matching strategies has its own advantage and  disadvantage. The MCMF strategy is parameter-free, has the best accuracy performance, but its time complexity is higher than its greedy counterpart; The greedy strategy is more time efficient, but it has additional parameter-traversal cost and its accuracy performance has a gap to the best value.

\section{Concluding remarks and future work}

The popularity bias is a ubiquitous problem confronted by common recommenders, and many research efforts were devoted to mitigate this problem and thus improve user experience. In fact, popularity bias is not welcome to not only users but also item-providers. While this problem of recommending a few popular items to a majority of users is usually regarded as a user-oriented problem, item-providers also suffer a lot from it and anchor their hope on recommender systems to give fairer exposure chance to different items, especially unpopular ones.

This work was devoted to solve the  problem of recommendation fairness, which is measured by the Gini coefficient of numbers of recommendation times of all items in the system.
The approach is to limit the allowed recommendation frequency of each item to be proportional to its degree. Although this approach is very effective and robust in significantly improving the recommendation fairness and thus better serves the purpose of item-providers, the following recommendation accuracy loss and the decrease of user experience cannot be ignored. To solve this subsequent problem, we proposed a heuristic strategy and an elaborate MCMF model to solve the user-item matching problem with constraints, both of which regain more than $100\%$ of the precision value of the baseline algorithm in regular recommendation context. Even compared with several state-of-the-art enhanced algorithms, the precision value of our P3-MCMF algorithm has no significant difference.
Another important advantage of our proposed P3-MCMF algorithm is that it is parameter-free and thus achieves this superior performance without the time cost of parameter optimization, while most existing enhanced algorithms have to traverse their intrinsic parameter to get the best performance.

\begin{figure*}[!htb]
\centering
  \subfigure[P3]{
    \includegraphics[width=2.5in]{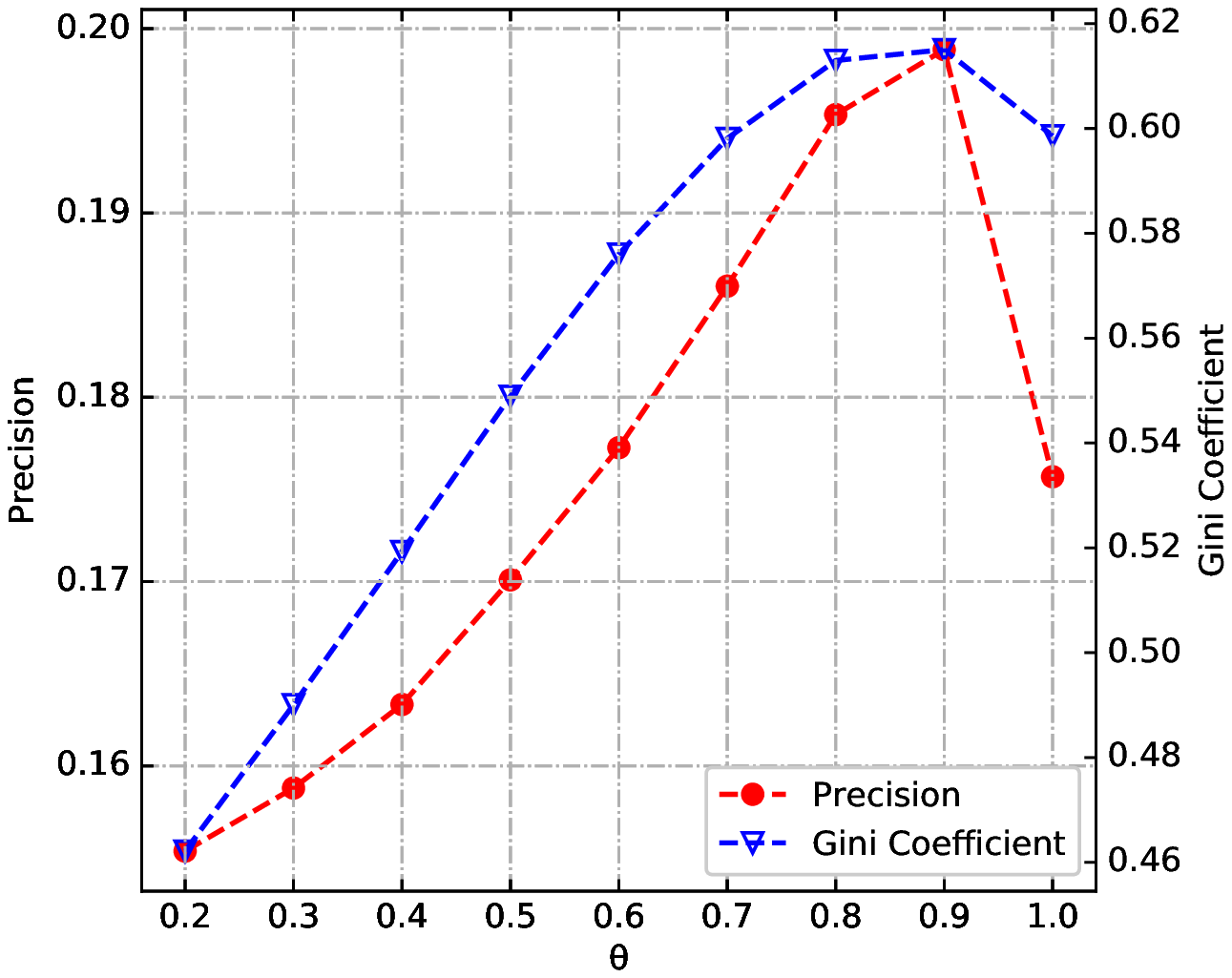}}
  \subfigure[RP3]{
    \includegraphics[width=2.5in]{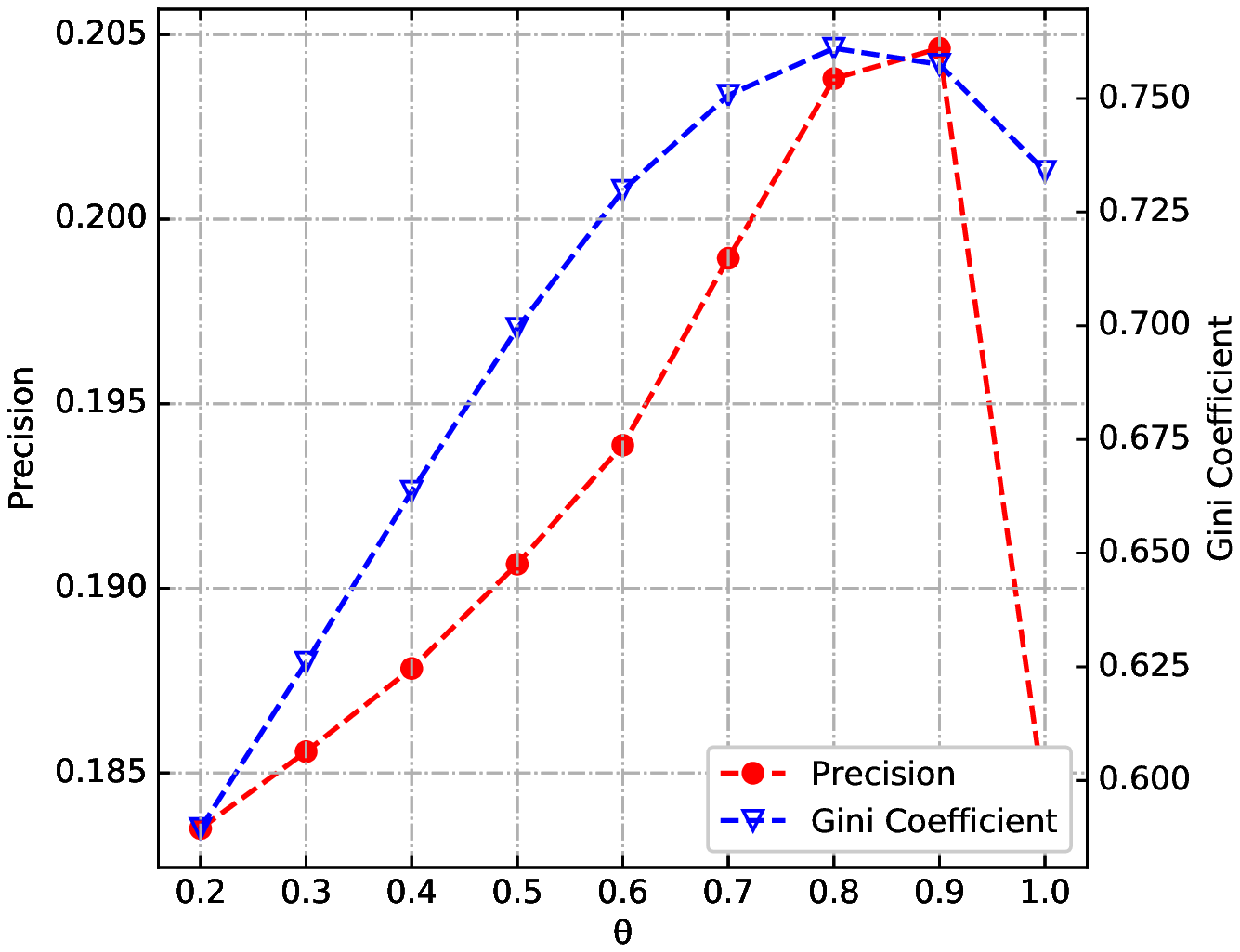}}
  \subfigure[HHP]{
    \includegraphics[width=2.5in]{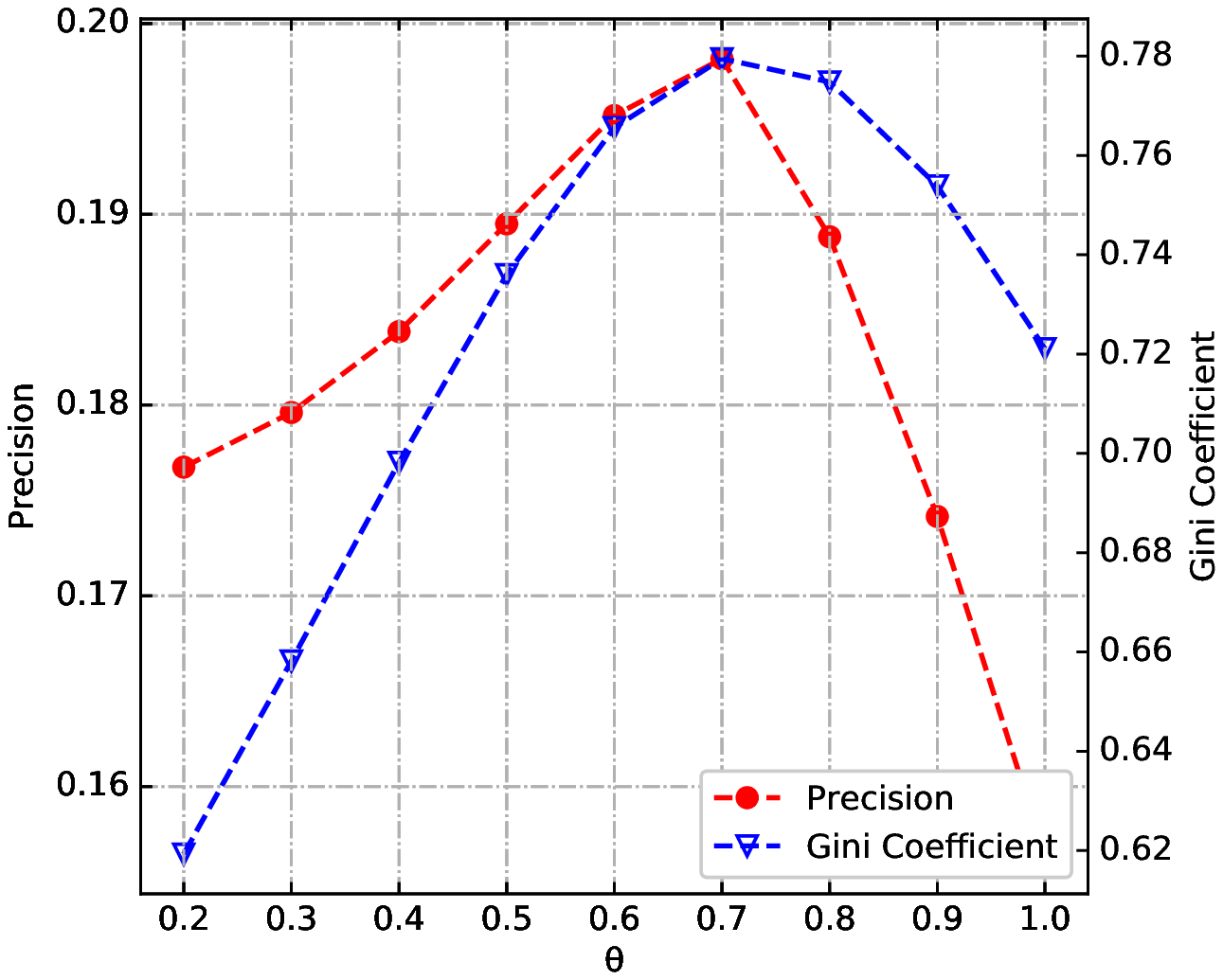}}
  \subfigure[BHC]{
    \includegraphics[width=2.5in]{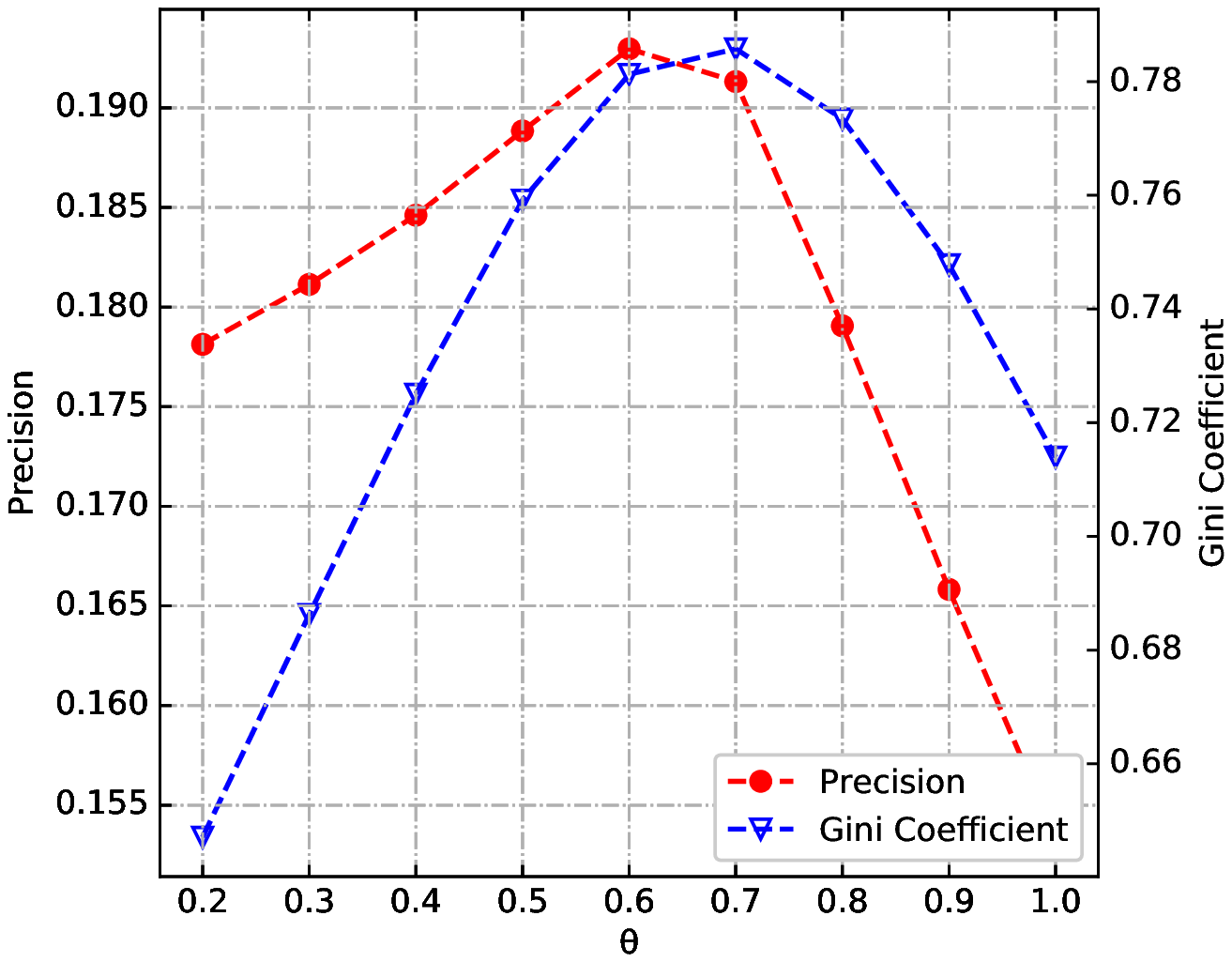}}
  \subfigure[BD]{
    \includegraphics[width=2.5in]{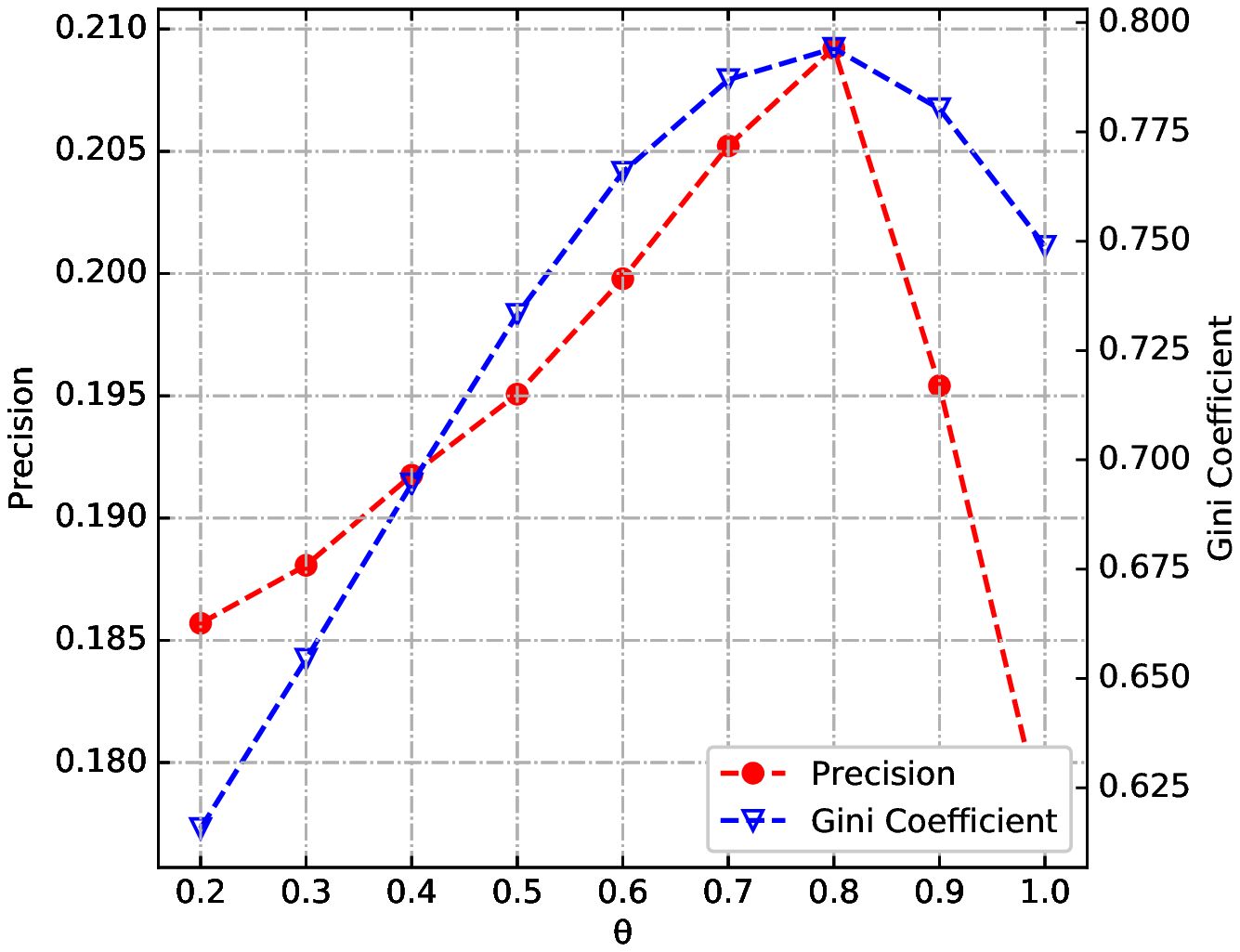}}
  \subfigure[PD]{
    \includegraphics[width=2.5in]{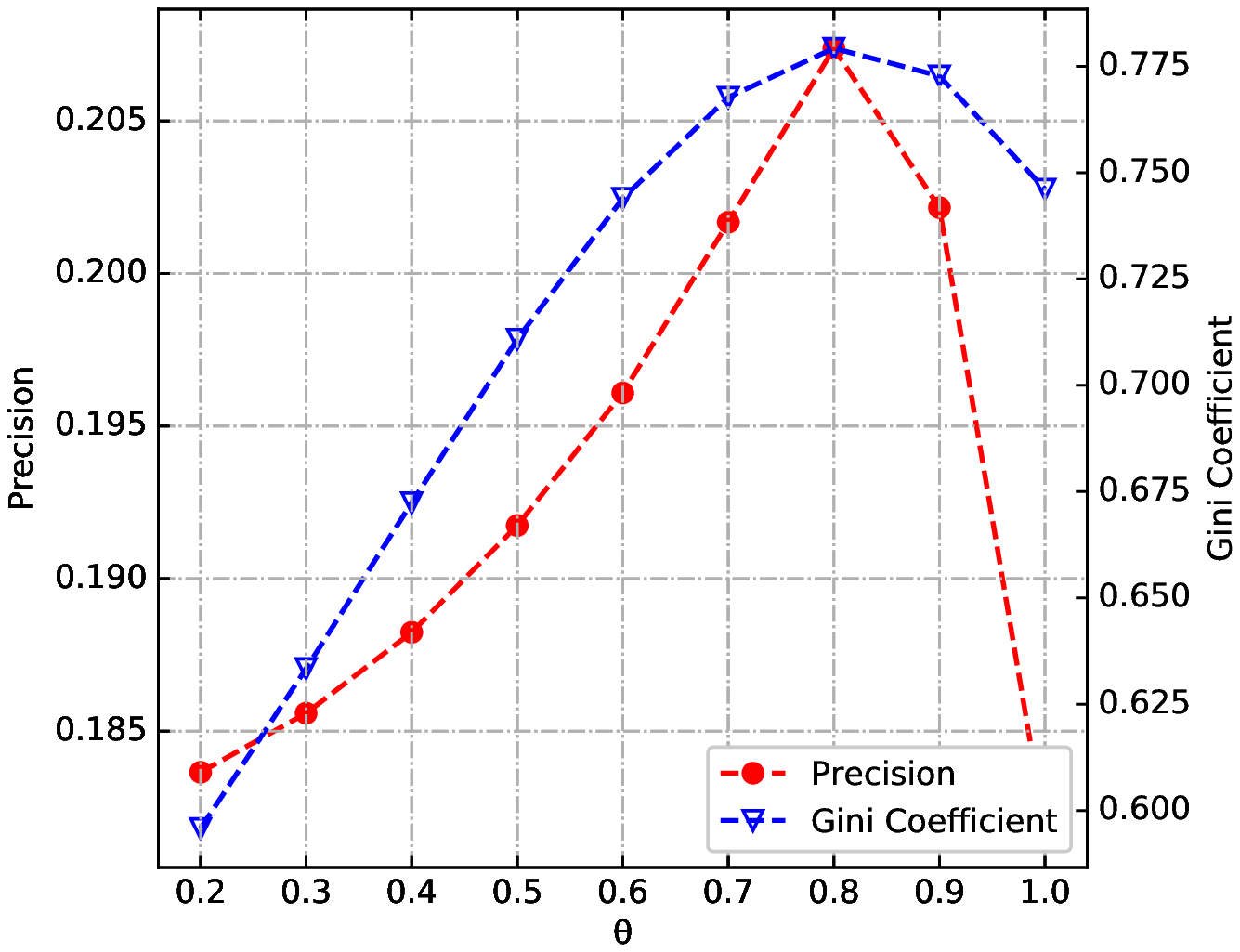}}

  \caption{The changes of the precision value and the Gini coefficient of recommendation scores in the recommendation lists of length 20, with the exponential parameter of the parameterized greedy strategy for P3 and five P3-enhanced algorithms on Movielens.}
\label{fig:peak}
\end{figure*}

In the greedy user-item matching strategy, by reducing the positive correlation between recommendation scores and user degrees via score normalization process, the recommendation accuracy and the distribution uniformity of recommendation scores are both improved. Then a natural question is raised, what is the relationship between these two measures? To answer this question, Fig. \ref{fig:peak}(a) plots the changes of the precision value and the Gini coefficient of recommendation scores in the recommendation lists of length $l = 20$, with the exponential parameter of the parameterized greedy strategy for the P3 algorithm on Movielens. We can see a very interesting phenomenon that, the two measures achieve their own optimal values at (almost) the same value of the exponential parameter. To see whether this is a common phenomenon in recommender systems or not, we regard the above-mentioned five P3-enhanced algorithms with optimal values of intrinsic parameters as baseline algorithms, and present the same plots in Fig. \ref{fig:peak}(b)-(f). This consistence between the peaks of the two measures still holds. Our next work is to explore the mechanism behind the consistence of these two peak values, and try to make use of it to further improve existing recommendation algorithms or design new recommendation models.

\section*{Acknowledgements}
The authors would like to express their gratitude to the editor and the two anonymous reviewers for their valuable suggestions that have greatly improved the quality of the paper.








\end{document}